# Observation of spin Nernst effect in Platinum

**Arnab Bose[1], Swapnil Bhuktare[1], Hanuman Singh[1], Venu Gopal Achanta [2] and Ashwin A. Tulapurkar*[1]**

*1 Dept. of Electrical Engineering, Indian Institute of Technology- Bombay, Mumbai 400076, India*
*2 Department of Condensed Matter Physics and Material Sciences, Tata Institute of Fundamental Research, Homi Bhabha Road, Mumbai 400005, India*

The spin direction of a nano-magnet can be efficiently manipulated by spin current injection. Several mechanisms are implemented to create spin current like electrical injection, spin Hall effect [1-6], spin Seebeck effect [7-8], spin pumping [9] and many more [10-11]. By virtue of spin Seebeck effect (SSE) pure spin current is generated in ferromagnet while temperature gradient is applied. In contrast heavy metals having large spin orbit coupling can convert charge current into pure spin current via the spin Hall effect (SHE). Spin current generated by these mechanisms can produce torque on nanomagnet which can be useful in memory and logic applications [12-15]. But the fundamental question: what happens if non magnet with large spin orbit coupling carries heat current, has not been studied experimentally. In this paper we report a new way of generation of spin current in heavy metal like Platinum (Pt) by applying temperature gradient which can be interpreted as spin Nernst Effect (SNE) [16-20]. We have demonstrated that non-magnetic Pt can convert heat current into pure spin current by virtue of SNE, which can be injected to nearby magnetic contact to obtain measurable voltage. We have used Pt Hall bar structure with ferromagnetic Ni detectors which allow us to compare the relative strength of SNE and SHE in the same sample.

In ordinary Hall effect [OHE] [21] while electric current is passed perpendicular to the applied magnetic field, Hall voltage is generated normal to the direction of both electric current and magnetic field since electron is deflected by the Lorentz force. Likewise if heat current is passed instead of charge current, open circuit voltage is developed normal to both heat current and applied field direction. This is known as Nernst-Ettingshausen Effect [NE] [22]. Interestingly when heavy metal (HM) like Pt carries electric current, up and down spins separate in opposite direction orthogonal to the direction of current flow (Figure 1.a) even without application of any external magnetic field. This is known as spin Hall effect (SHE). SHE was theoretically predicted long back [1-2] and experimentally observed in past few years [3-5]. This effect arises due to the coupling between electrons spin angular momentum and its orbital motion, which originates from the relativistic Dirac equation. Heavy metals like Pt are good candidates for observation of SHE due to the large spin-orbit coupling. There are two possible mechanisms of SHE. It can arise from the internal band structure of a material where scattering plays a minor role (intrinsic SHE) [23] or it can arise from spin dependent scattering of electrons with the impurities present in the material (extrinsic SHE) [24]. Now question is: if heavy metal like Pt is set between two temperature baths, can it generate pure spin current. If thermal gradient is created in a metal, in open circuit condition Seebeck voltage is generated across it. In this condition, internally electrons can flow maintaining net charge current equal to zero. If we look at the energy resolved electron current, the electrons below the Fermi level (cold electrons) flow along the temperature gradient, while electrons above the Fermi level (hot electrons) flow in opposite direction. As shown in fig 1.b, the hot electrons are scattered sideways due to the spin orbit interaction (SOI), resulting in spin current $J_{spin1}$. Similarly cold electrons give rise to spin current $J_{spin2}$, which is opposite to $J_{spin1}$ as the cold electrons flow opposite to hot electrons (Fig 1.b). If the SOI scattering rate is the same for hot and cold electrons, the net spin current wold be zero. However, if the scattering rates are different ($J_{spin1} \neq J_{spin2}$), a non-zero spin current can be created in heavy metals perpendicular to the flow of heat current (Fig 1.c). This effect can be interpreted as spin Nernst effect or thermally driven spin Hall effect. It is to be noted that metals with large spin orbit coupling is not a sufficient condition to observe SNE, the scattering should have large energy dependence at Fermi level. Since last few years there were predictions [17-18] of SNE but it was lacking proper experimental evidence. We have employed multi-terminal Ni/Pt junctions to compare the strength and relative sign of SHE and SNE of Pt. Our result is consistent with very recent report of SNE [19-20]. With the discovery of SNE two separate fields viz. spin-orbitronics [1-6] and spin-calortitronics [25-30] can be merged together to form spin-orbito-caloritronics (Fig 1.d).

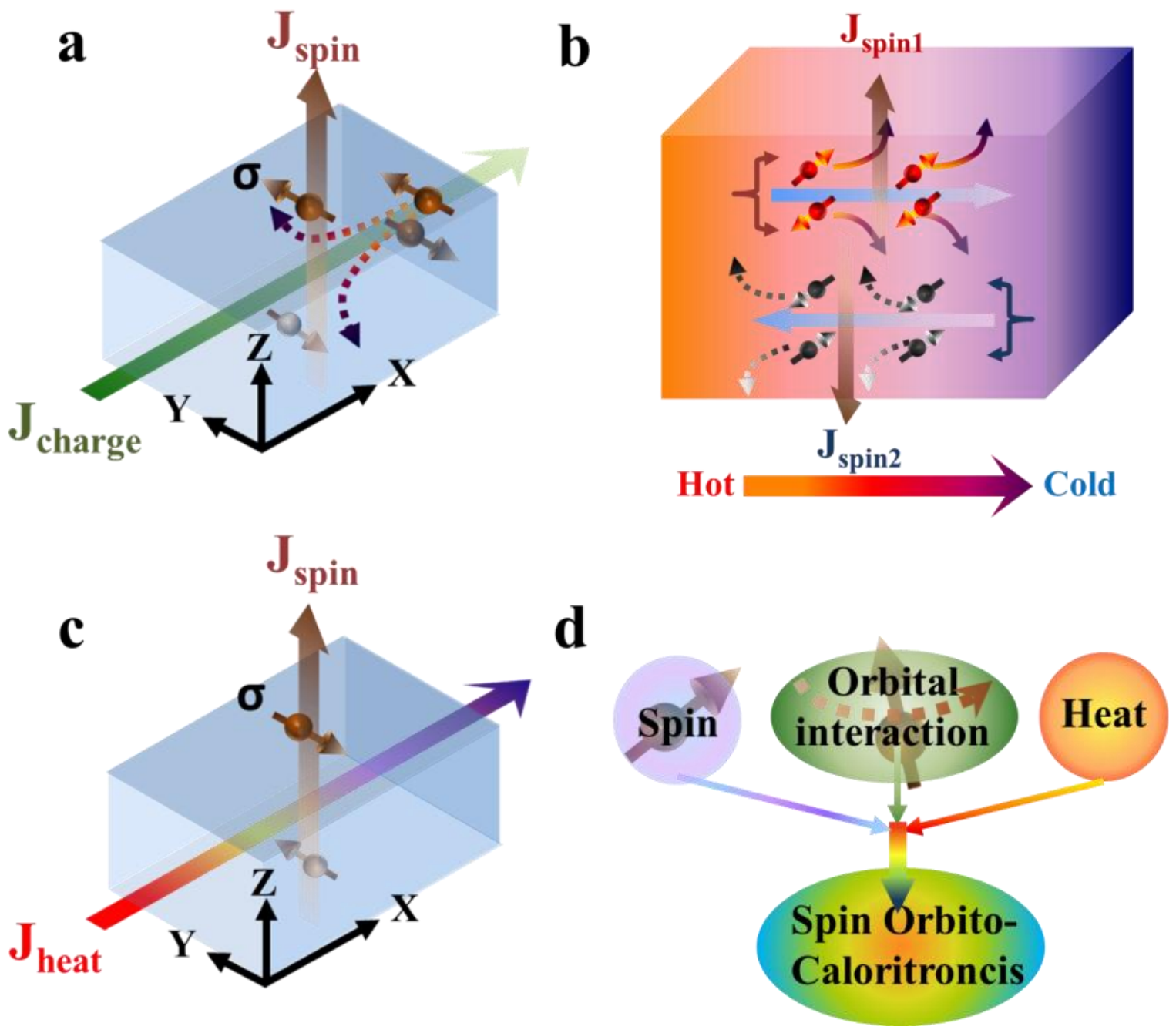

**Figure1 | Conception of spin Hall effect (SHE) and spin Nernst effect (SNE)**. **a**, Schematic diagram of SHE (charge current to spin current conversion in heavy metals (HM)). The electrons flow along –X direction, spins oriented along +Y (-Y) are accumulated at top (bottom) surface of HM. **b**, Microscopic origin of SNE. When temperature gradient is applied along HM, hot electrons (shown by red colour) and cold electrons (shown by deep blue colour) flow in opposite directions. The spin current produced by hot ($J_{spin1}$) and cold ($J_{spin2}$) electrons is in opposite direction. A non-zero net spin current ($J_{spin1}$-$J_{spin2}$) would flow transverse to thermal gradient if the spin-orbit scattering rate is different for hot and cold electrons. **c**. Conceptual picture of SNE (heat current to spin current conversion) which is equivalent to figure b when scattering rate for hot and cold electrons are differnt. **d**, New emerging field of spin-orbito-caloritronics which considers interplay of electrons spin and orbital interaction in presence of temperature gradient.

Experimental procedure is described in figure 2. Figure 2.a shows the coloured Scanning Electron Microscopic (SEM) image of fabricated device and figure 2.b, 2.c show the schematic diagrams of SNE and SHE experiments respectively. A Hall cross-bar structure (blue cross in fig 2.a) is prepared by electron beam lithography (EBL) and sputter deposition of 10 nm thick Pt. Then Ni lines (thickness 10 nm) are deposited by EBL, sputtering and lift off technique (green lines as shown in figure 2.a). Before deposition of Ni, top surface of Pt is cleaned in-situ by Argon ions to make transparent contact. Final contacts are made by Ti/Au (shown as yellow colour in fig 2.a). All depositions are done at base vacuum better than 8E-8 Torr. Separation of Pt line and Ni line is approximately 2 μm (Fig 2.a). Typical length and width of Hall bar is 40 μm and 6 μm.

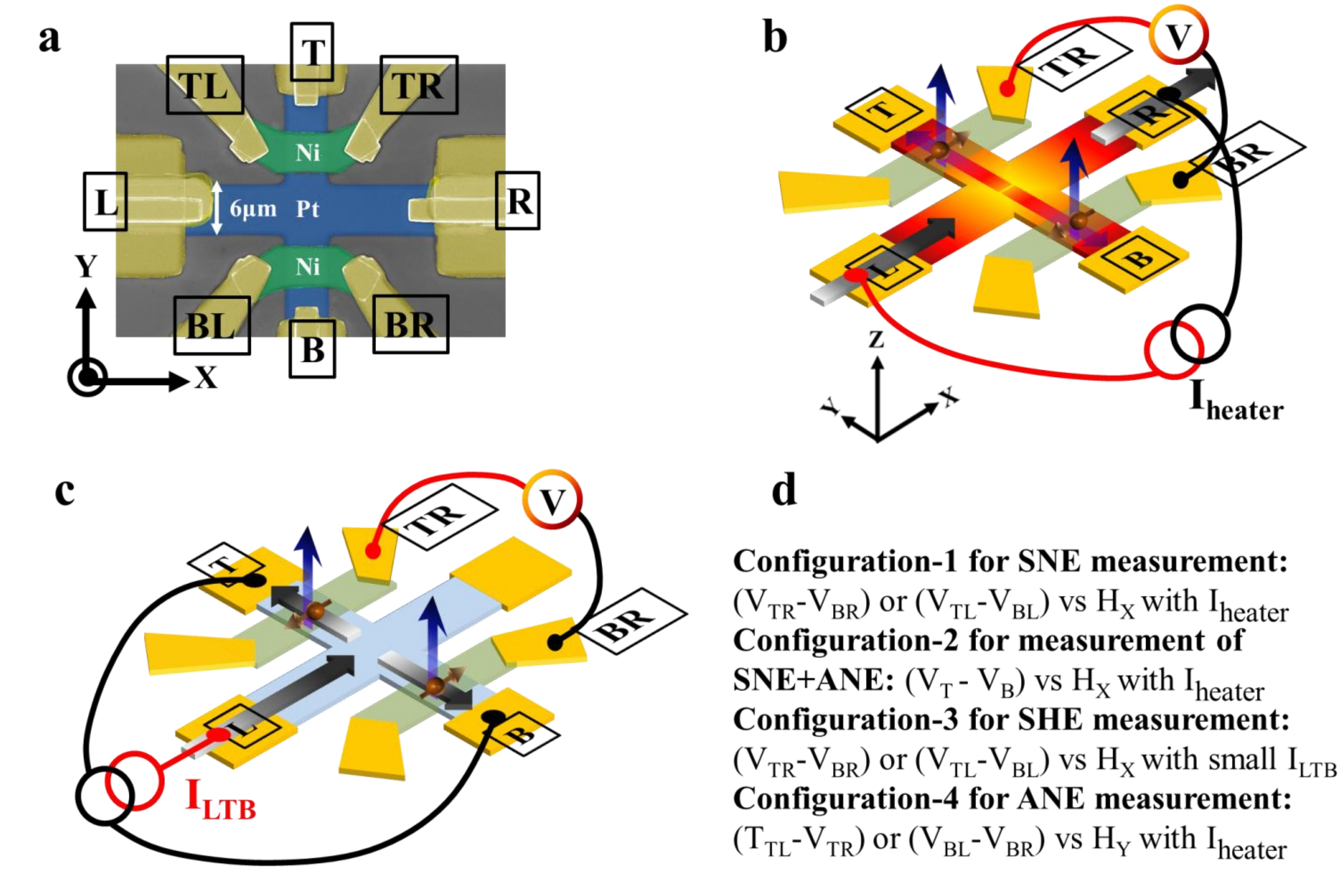

**Figure 2| Experimental description. a**, Coloured SEM image of the fabricated device. Hall bar (blue colour) is made of Pt. Two Ni lines (green colour) are deposited on Pt. **b**, schematic representation of SNE experiment. Electric current ($I_{LR}$) flows through Pt line between two terminals 'L' and 'R' along X axis and Joule heat is generated at centre of Hall bar and flows towards Ni line along Y axis. $V_{SNE}$ is measured between two different Ni contacts sweeping field along X axis). **c**, schematic description of SHE experiment. Charge current flows in Pt Hall bar such that two Pt lines below Ni contacts carry current in opposite direction. $V_{SHE}$ is measured between contacts of two different Ni terminals as a function of field ($H_x$). For both the experiments of SHE and SNE spins are injected from Pt to Ni along Z axis (indicated as blue arrow pointing out of plane direction). So if up spins (+X orientated) are accumulated in top Ni line then down spins (- X orientated) will be accumulated in bottom Ni line since direction of heat current (in SNE experiment) and electric current (in SHE experiment) is opposite. **d**, Summary of the experiments: SNE, SNE+ANE,ANE and SHE.

For SNE experiment large dc electric current ($I_{heater}$) is passed along heater line (between terminals 'L' and 'R' of Pt). While Pt carries current, centre of the Hall bar is heated due to Joule heating and heat flows towards the Ni lines (Fig 2.b). Hence temperature gradient is created along Ni/Pt bilayer along ±Y direction. Pt converts heat current into spin current by SNE which is injected in to Ni. Since Ni is fabricated on top of Pt, spatial direction of spin flow from Pt to Ni is out of plane (+Z axis, shown with blue arrows in figure 2.b,c) but heat flow direction is opposite (along ±Y) in two different Ni wires (Fig 2.b). Hence spins of opposite polarity (along the direction of ±X) are accumulated into Ni at non equilibrium condition. If magnetization of the Ni contacts point along same direction (which happens for magnetic field greater than coercivity), we expect to see differential voltage step between the contacts of top and bottom Ni lines ($V_{TR}$-$V_{BR}$ or $V_{TL}$-$V_{BL}$) as a function of magnetic field sweep along X axis as shown in figure 2.b. The signal (kinks) near zero field is due to the domains present in the Ni lines and hence we focus on the voltage signal for large magnetic fields. In this configuration we'll get only contribution from SNE. All figures shown in figure 3 are obtained after averaging positive and negative heater current so that we get pure signal generated by heating effect and any sort of electrical voltage will be cancelled. We measured clear voltage step of 120 nV (approx.) between top and bottom Ni contact (Fig 3.a) while applied heater current ($I_{heater-average}$) is 15 mA. Further we verified that this observed voltage step is proportional to square of average heater current and voltage step sign is independent of the polarity of applied current (see Fig S3.2 in supplementary information for more details). In this configuration-1 we can only observe contribution from SNE as mentioned in Fig 2.d. Differential measurement of voltage enables us to detect SNE voltage step with higher sensitivity since it reduces background voltage significantly. For maximum heater current ($I_{LR}$=±15mA) estimated temperature gradient in Pt/Ni bilayer is approximately 8.5 K/μm (see Section S1 in supplementary information). During the experiment resistance of Pt ($R_{LR}$) heater line and Ni line ($R_{LR}$ (top) and $R_{LR}$ (bottom), (see figure 2.a) are monitored which provides on-chip temperature calibration. Further it is compared with COMSOL simulation. There could be unintentional temperature gradient along out of plane direction (Z axis) which turns out to be quite small as compared to the in-plane temperature gradient (see S1 in supplementary material). We have performed additional experiments to observe the impact of Z direction temperature gradient on SNE. Out of plane temperature gradient can contribute to anomalous Nernst voltage [29-30] according to the relation $\nabla V_{ANE}(y) \; \alpha \; \left(\hat{M}(x)\times\nabla T(z)\right)$ where M, T, V represent magnetization, temperature and voltage respectively. We can refer configuration-2 to compare voltage step due to combination of SNE and ANE. In configuration-2 voltage is measured between top ('T') and bottom ('B') contact (Fig 2.a) with external field sweep along ±X. In this configuration observed voltage step is approximately 150 nV which is quite comparable to voltage step observed in configuration-1 ($\Delta V_{SNE}$~120 nV) (see Fig 3.a,b). In this way we can argue that ANE signal is quite small compared to SNE signal. Even in the same device geometry we can measure the pure signal of ANE by measuring voltage along same Ni line ($V_{TL}$-$V_{TR}$ or $V_{BL}$-$V_{BR}$) as a function of external field sweep along Y. This is shown configuration-4 (see Fig 2.d). In configuration-4 magnetic field is swept along Y which is orthogonal to the direction of spin accumulation (±X). Hence no SNE signal is generated but ANE voltage can be measured along same Ni line due to the relation: $\nabla V_{ANE}(x) \; \alpha \; \left(\hat{M}(y)\times\nabla T(z)\right)$, where voltage drop is in Ni (along X), magnetization is swept along Y and thermal gradient is along Z. Even in this configuration ANE voltage is quite negligible (~20nV) compared to SNE signal (~120 nV) (see supplementary information S3.1 for more details). Ni-Pt junction size is typically (5x5) $\mu m^2$ (Fig1.a). Experimental configuration 1-3 involves measurement of differential voltage which is highly sensitive compared to configuration-4. Additional experiments are done to rule out possibility of other spurious effects. In configuration-1 it is shown that SNE voltage step is observed when magnetic field is parallel to the direction of spin accumulation (X axis). In contrast when voltage is measured between two different Ni terminals ($V_{TL}$-$V_{BL}$ or $V_{TR}$-$V_{BR}$) sweeping magnetic field along Y we notice that voltage step disappears (Fig 3.d). It is due to spin polarization direction (along ±X) is orthogonal to the field sweep direction (±Y). Finally when Pt is substituted by Al (low spin orbit coupling) measured SNE voltage is significantly reduced (Fig 3.e). More experimental results are shown with details of power variation and sample variation in supplementary information [S3]. It indicates that we have consistently observed finite step in SNE voltage with same sign in all experiments which cannot be ascribed by other spurious effects like asymmetry in differential magnetic thermopower, non saturating domain activity difference in different Ni contacts and even magnetic proximity effect (see section S4 in supplementary information). All of the experimental observation supports the fact that heat current can be converted into spin current by SNE in nonmagnetic Platinum.

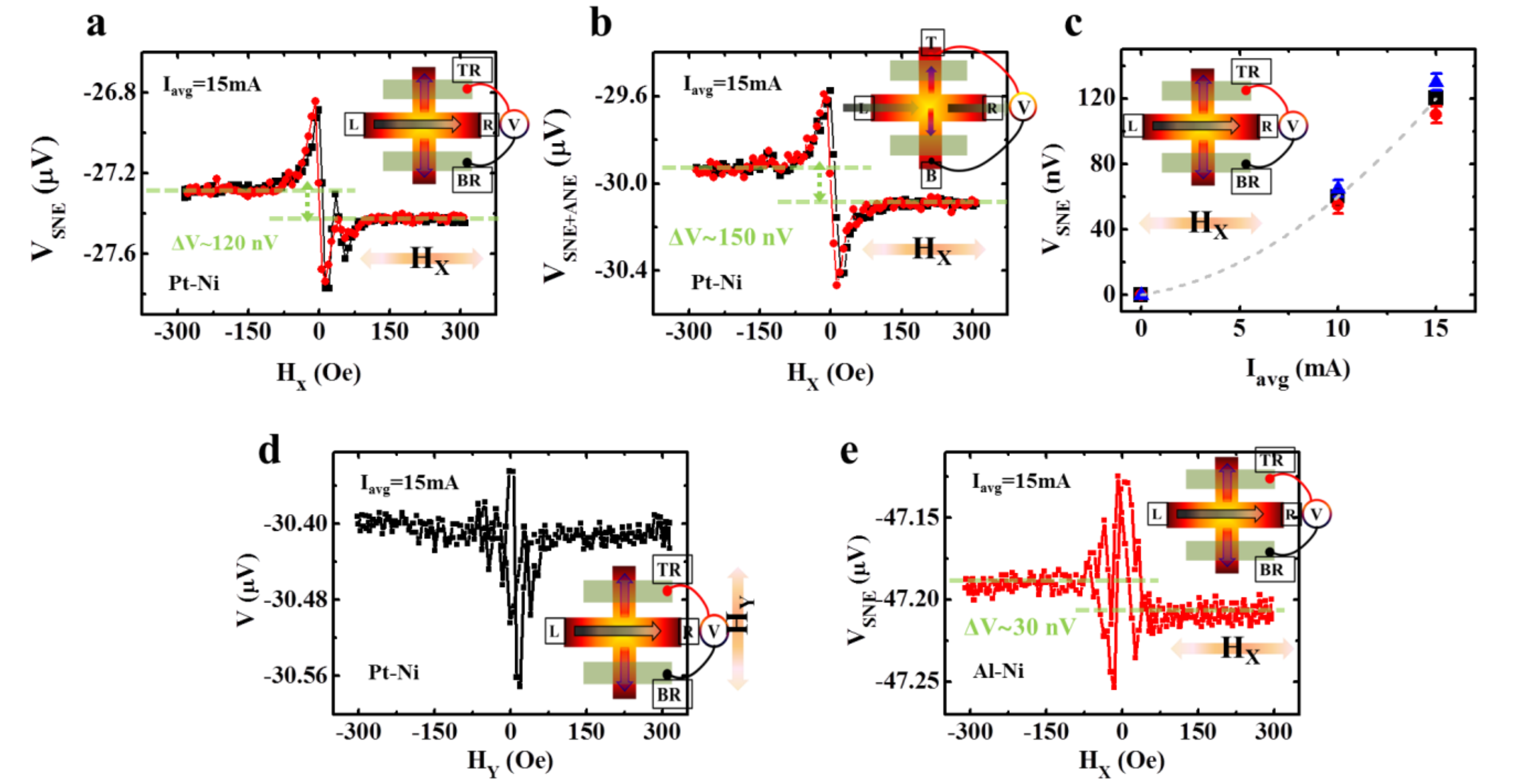

**Figure 3|Experimental result of spin Nernst effect**. **a**, $V_{SNE}$ measured between two different Ni lines for 15 mA average current. **b**, Measured ($V_{SNE}$+$V_{ANE}$) voltage as a function of external field sweep along X when voltage is measured between top (T) and bottom (B) contacts. **c**, Step of SNE voltage as function of average heater current. **d**, Measured voltage between different Ni lines as a function of field sweep along Y. **e**, measured $V_{SNE}$ voltage when Pt is replaced by Al.

Now we measure SHE in the same device (configuration-3 in figure 2.d). Electric current is passed between terminals L-T ($I_{LT}$) and L-B ($I_{LB}$) as shown in figure 2.c. In this arrangement current flow direction in each of Pt/Ni is opposite (±Y). While Pt conducts electric current along ±Y direction it injects spins with orientation ± X direction into Ni by SHE. This is also similar to earlier situation where spins of different polarity are accumulated in Ni detector (Fig 2.b,c). Only difference is that here spin current is generated in Pt by SHE whereas in previous case spin current was created by SNE. So we measure voltage between two different Ni lines (top and bottom Ni lines as shown in fig 2.c) as a function of external field sweep along X direction. We observe clear step in voltage ($\Delta V_{SHE}$) for saturating magnetic field and step-sign reverses with changing the polarity of electric current flow direction (Fig 4.a, b). In our control experiment we replace Pt by Al (low spin orbit coupling) which also does not show any step in measured voltage (Fig 4.d). So it clearly indicates that Pt converts charge current into spin current and injects into Ni which floats to measurable voltage as a function of external field (along X direction). It is notable that additional peaks or dips are observed in measured voltage near the zero magnetic field in figure 2 and 3. It can occur due to combination of planar Nernst and thermal AMR (planar Hall and AMR) effect in the experiment of SNE (SHE) (Fig 2.b,c). PNE (PHE) corresponds to generation of transverse voltage (along X) while it carries heat (electric) current (along Y) [29,30]. In ideal condition PNE (PHE) generated in each of the Ni branch should get cancelled but there is always some asymmetry in device fabrication and they do not cancel each other resulting kinks in near the field where domains of Ni rotate. According to the relation $\nabla V_{PNE(PHE)}\ \alpha\ \left(\hat{M}\times\hat{M}\times\nabla T\right)$, PNE (PHE) voltage shows sin2θ dependence while magnetic field is rotated with respect to the current (heat or charge) flow direction. So PHE and PNE do not exhibit any step in measured voltage but they can manifest as kinks due to rotation of magnetic domain while field is swept [29,30]. Similar arguments apply to thermal AMR and electrical AMR ($V_{AMR}$ α $\cos^2\theta$), and these effects also do not result in any step in the voltage [29, 30]. Hence steps observed while voltage is measured between different Ni contacts have to come from SNE or SHE. We have tested different sets of samples which reproduced similar behaviour of voltage step for SHE and SNE. The error bars shown in fig 3c include the sample variation (see section S3.2 in supplementary information). Unlike the previous study of SHE [3,4] where non-local detection method was used, we implement local detection method with multi-terminal device structure which further reduces the background (ANE/SNE) significantly. Hence this local detection method enables us to observe SNE directly and compare with SHE.

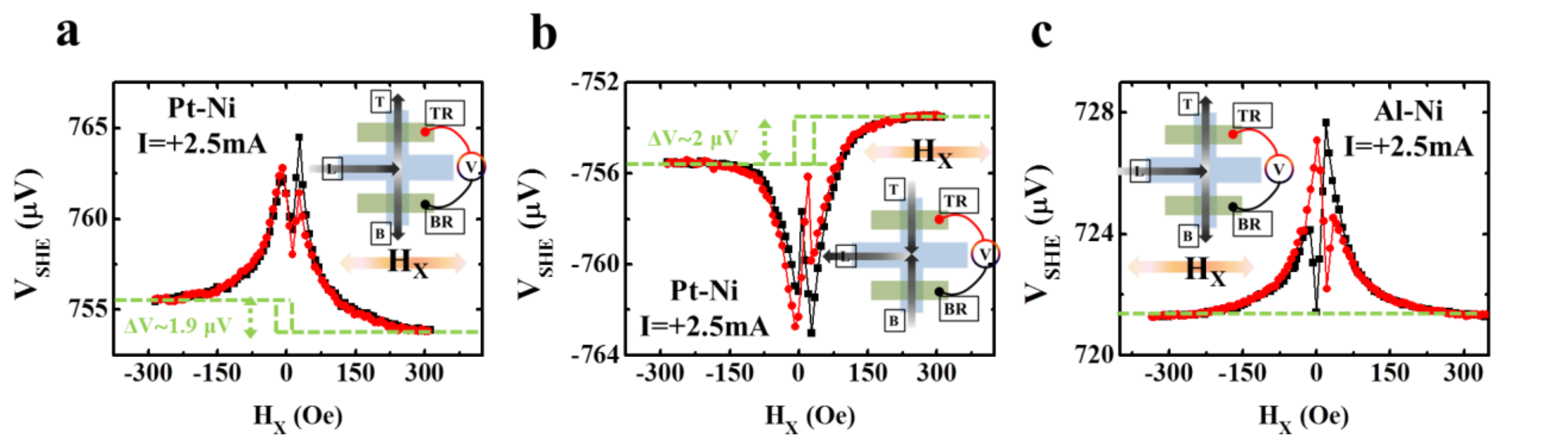

**Figure 4| Local detection of spin Hall effect (SHE).** **a**, **b** Voltage step (ΔV) sign of SHE changes with reversing the polarity of passed current (±2.5 mA) in Pt/Ni while field is swept along X axis. **c**, when Pt is replaced by Al step becomes negligible.

Now we compare the results of spin Hall effect and spin Nernst effect. We extract spin Hall angle ($\theta_{SH}$) of Pt from the experiment shown in figure 4. We observed approximately 1.9 μV step ($\Delta V_{SHE}$) when each of Ni/Pt branch carries 2.5 mA current. We estimated how much spin current density is needed to obtain voltage step of 1.9μV. Considering electrical conductivities of Ni and Pt 2(ohm-μm)$^{-1}$ and 6(ohm-μm)$^{-1}$, spin flip length of Ni 3 nm, spin polarization of Ni ($\beta_{FM}$) 0.26 we obtain spin Hall angle of Pt ($\theta_{SH}$) 0.07 which is very close to other reported values [1-5]. In contrast to SHE when Pt carries thermal gradient of order 8.5 K/μm (see S1 in supplementary materials for temperature calibration) we observe step ($\Delta V_{SNE}$) of 120 nV between two Ni lines from which we can estimate that Pt converts 8.5 K/μm temperature gradient into pure spin current density of $1.43\text{x}10^8$ A/m$^2$. To get dimensionless spin Nernst angle ($\theta_{SN}$) we have to convert thermal gradient into equivalent charge current density [19]. So the equivalent charge current density driven by thermal gradient in short circuit condition can be represented by $J^{Q'} = \sigma Q\left(-\frac{dT}{dY}\right)$ where σ, Q, $\frac{dT}{dY}$ represent the electrical conductivity, Seebeck coefficient and thermal gradient along Pt. Assuming Seebeck coefficient of Pt to be -6.25 μV/K at 340K [31] spin Nernst angle ($\theta_{SN}=J_s/J^{Q'}$) turns out to be close to -0.45. In this calculation we have neglected the interfacial resistance effect [see supplementary information]. Magnitude of Spin Nernst angle can slightly vary based on the thermal interfacial resistance and values of different parameters used. Magnitude of reported value of spin Nernst efficiency is more (-0.45) than spin Hall efficiency (0.07) but opposite in sign. It is to be noted that equivalent charge current density in short circuit condition may not be directly related to generation of spin current density from thermal gradient which depends on asymmetry of spin scattering around Fermi energy level ($E_F$)as argued earlier (Fig 1). On this argument we may have some material with finite Seebeck coefficient and spin orbit coupling but exactly symmetric scattering rate around EF will not convert heater current to spin current by SNE. However this definition of dimensionless spin Nernst angle will be useful to compare various experimental results.

In conclusion we have demonstrated that heavy metal, Pt can convert heat current into spin current by spin Nernst effect. We comparatively studied spin Hall effect and spin Nernst effect in Ni/Pt multi-terminal bi layer heterostructure which reveals that spin Hall angle and spin Nernst angle in Pt are opposite in polarity. Our method to detect spin current by measuring differential voltage is very sensitive. Generation of spin current by temperature gradient in heavy metals involves interplay of spin, orbit-coupling and thermal gradient which opens a new field of spin orbito-caloritronics.

## AUTHOR INFORMATION

### Corresponding Author

* Email: ashwin@ee.iitb.ac.in

### Author Contributions

The device fabrication and measurements were carried out by AB. SB and HS helped AB in experiments. AB analysed the data and wrote manuscript with help from AT. VGA gave inputs and experimental support for local heating experiments. AT supervised the project. All authors contributed to this work and commented on this paper.

Note: The authors declare no financial interests.

## ACKNOWLEDGMENT

We would like to acknowledge the support of Centre of Excellence in Nanoelectronics (CEN) at IIT-Bombay Nanofabrication facility (IITBNF), Indian Institute of Technology Bombay, Mumbai, India.

# Supplementary Information

**S1. Temperature Calibration**

**S2. Estimation of spin Hall angle and spin Nernst angle**

**S3. More experimental data**

**S3.1 Comparison of measured voltage from various terminals**

**S3.2 Details of power variation with positive and negative heater current**

**S4. Effect of current spreading near Ni/Pt bilayer**

**S5. Proximity effect and other possible source of voltage step vs SNE signal**

**S1. Temperature calibration:**

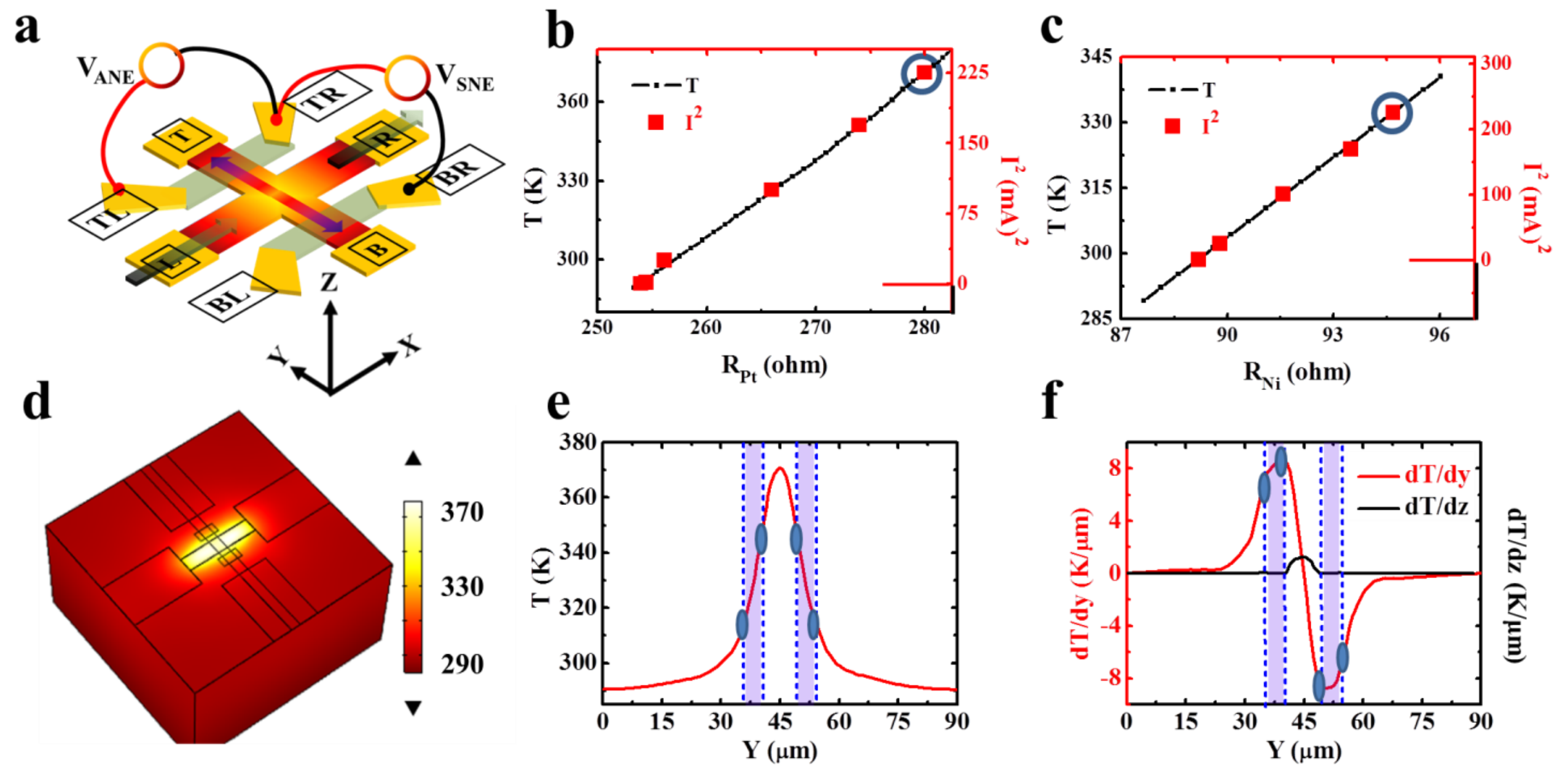

**Figure S1**| Temperature calibration. **a**, Experimental device set up. **b**, **c** estimation of average temperature of Pt and Ni line from its resistance value as a function of temperature. **d**, 3D temperature profile of the device. **e,** estimated temperature profile along Y axis from COMSOL simulation. **f**, Estimated temperature gradient along Y axis. Blue stripes indicate the position of Ni line.

As shown in figure S1.a two Ni lines are fabricated on Pt Hall bar. While current flows in Pt heater line between terminal ‘L’ and ‘R’ (along X direction) centre of Hall bar is heated up and heat flows towards Ni terminals (along Y). During the experiment resistance of Pt (between terminals ‘L’ and ‘R’) is monitored which shows linearly increasing with square of applied current ($I_{LR}$) (right Y axis of Fig S1.b). Resistance of Pt (between ‘L’ and ‘R’) as a function of temperature is also measured (left Y axis of Fig S1.b). From this we obtain the temperature of hot spot created by Joule heating. For ±15 mA applied current ($I_{LR}$) maximum temperature reaches close to 370 K (left axis of Fig S1.b) which is close to the predicted value from COMSOL simulation (Fig S1.e). Similarly resistance of each Ni line (between terminals {‘TL’ and ‘TR’} or {‘BL’ and ‘BR’}) is monitored while passing current in Pt heater line ($I_{LR}$, between terminals ‘L’ and ‘R’) (right Y axis of Fig S1.c). Resistance of same Ni line is also measured as a function of temperature from (left Y axis of Fig S1.c). From this measurement we get average temperature of Ni line to be approximately 332 K which is close to the predicted temperature value by COMSOL simulation (Fig S1.e). From figure S1.e we see that approximately 40 K temperature difference along Ni/Pt bilayer of width 5 μm (along Y axis). We can assume that average in-plane temperature gradient (dT/dy) in Ni/Pt bilayer is roughly 8.5 K/μm (fig S1.f). Now we try to find the out of plane temperature gradient in Ni experimentally. For this we measure Anomalous Nernst (ANE) $[\nabla V_{ANE}(x)\,\alpha\left(\hat{M}(y)\times\nabla T(z)\right)]$ voltage at same Ni terminal (either between {‘TL’ and ‘TR’} or between {‘BL’ and ‘BR’}) as shown in the figure S1.a sweeping external field along Y axis. Figure 3.e in the main manuscript shows the measured ANE voltage step (~20 nV) corresponding to ±15 mA heater current ($I_{LR}$)). From the previous reports [30] assuming ANE coefficient to be 2.6 μV/K we calculate the average out of plane temperature gradient in Ni to be roughly 1 mK/μm which further exponentially decays in Ni along Y axis as we go away from hot spot. This value is also comparable from COMSOL prediction (Fig S1.f). Ni-Pt junction size is typically 5μm x 5μm.

**S2. Estimation of spin Hall angle and spin Nernst angle**

We estimated voltage considering Valvet-Fert equation [S1]:

Let's consider Valvet –Fert equation for electrons spins in magnet:

$$\frac{\partial^2 \Delta\mu}{\partial z^2} = \frac{\Delta\mu}{l_{sf}^2} \qquad (1)$$

where $\Delta\mu = \bar{\mu}_+ - \bar{\mu}_+$ is the difference between electrochemical potential for up (+) spin and down spin (-), $l_{sf}$ is the spin flip length of ferromagnet.

Now we consider conservation of charge current in the system: $\frac{\partial J_{ch}}{\partial z} = 0 = \sigma_+ \frac{\partial^2 \bar{\mu}_+}{\partial z^2} + \sigma_- \frac{\partial^2 \bar{\mu}_-}{\partial z^2}$ where $J_{ch}$ is the charge current and $\sigma_+$ ($\sigma_-$) is the conductivity of electron in up channel and down channel. We can rewrite it in following way:

$$\frac{\partial^2}{\partial z^2}\left(\sigma_+\bar{\mu}_+ + \sigma_-\bar{\mu}_-\right) = 0 \qquad (2)$$

Solving equation (1) and (2) we get following expressions:

$$\Delta\mu = \bar{\mu}_+ - \bar{\mu}_- = 2\left(K_2 e^{z/l_{sf}} + K_3 e^{-z/l_{sf}}\right) \qquad (3)$$

$$\bar{\mu} = \frac{\bar{\mu}_+ + \bar{\mu}_-}{2} = Je\rho\, z + K_1 + \beta\left(K_2 e^{z/l_{sf}} + K_3 e^{-z/l_{sf}}\right) \qquad (4)$$

$$J^{spin} = J_+ - J_- = -\beta J + \frac{1}{e\rho^* l_{sf}}\left(K_2 e^{z/l_{sf}} - K_3 e^{-z/l_{sf}}\right),\ \rho^* = \frac{1}{\sigma(1-\beta^2)} \qquad (5)$$

Where, J+ (J-) is the up (down) spin current density, $K_1$, $K_2$, $K_3$ are constants, $\beta = \frac{\sigma_{\text{MAJORITY}} - \sigma_{\text{minority}}}{\sigma_{\text{MAJORITY}} + \sigma_{\text{minority}}}$ is spin polarization in ferromagnet.

At Pt/Ni interface (Z=0) there is no charge current but Pt injects spin current into ferromagnetic Ni contact.

So $J_{ch}$ (z=0)=0 but $J^{spin}$ (z=0)≠0.

From equation (4) we get: $\bar{\mu} = K_1 + \beta(K_3 e^{-z/l_{sf}}) = \beta(K_3 e^{-z/l_{sf}})$, assuming $\bar{\mu}(z=\infty) = 0$

From equation (3) and (5): $\Delta\mu = 2(K_3 e^{-z/l_{sf}})$, $J^{spin} = J_+ - J_- = \frac{1}{e\rho^* l_{sf}}(-K_3 e^{-z/l_{sf}}) \Rightarrow K_3 = -e\rho^* l_{sf}\, J^{spin}(0)$.

From this equation we can write the expression of voltage generated in Ni while absorbed spin current:

$$\bar{\mu}(z=0) - \bar{\mu}(z=\infty) -= \beta K_3 = -\beta e \rho^* l_{sf}\, J^{spin}(0)$$

$$J^{spin} = \theta_{SH} J_{ch} \qquad \text{For spin Hall effect}$$

$$= \theta_{SN} J^{Q'} = \theta_{SN} \sigma Q \left( -\frac{dT}{dY} \right) \qquad \text{For spin Nernst effect}$$

Thickness of Pt and Ni is 10nm. Conductivity of Pt and Ni=6x10$^6$ and 2x10$^6$ (in SI unit) respectively. Assuming β=0.26 and $l_{sf}$=3 nm we get the magnitude of step to be approximately 1.95 μV while 2.5 mA current is passed in each of Ni/Pt branch (Figure 2.c in main paper) which corresponds to spin Hall angle of Pt 0.07. It implies that 2.4x10$^9$ A/m$^2$ spin current density is injected from Pt to Ni by SHE. This result is quite close to the previous experimental values [3-5]. Now we want to quantify spin Nernst angle comparing the data of SHE. As shown in SNE experiment (Figure 3.a in the main paper) we observed approximately 120 nV step while in-plane temperature gradient is created in Pt (8.5K/μm) along Y axis. From the above expressions and material parameters we can get that an estimate that 1.43x10$^8$A/m$^2$ spin current has to be injected from Pt to Ni to obtain 120 nV of voltage step. So 1.43x10$^8$A/m$^2$ spin current density is converted by Pt by SNE. From this we calculate that spin Nernst coefficient is approximately -0.45. It is to be noted that sign of spin Hall angle and spin Nernst angle is different. The exact value of $\theta_{SH}$ and $\theta_{SN}$ depends on the various material parameters. However in these calculations we have neglected interfacial resistance effect. If interfacial heat resistance is different than the interfacial charge resistance we may obtain different values of spin Nernst angle.

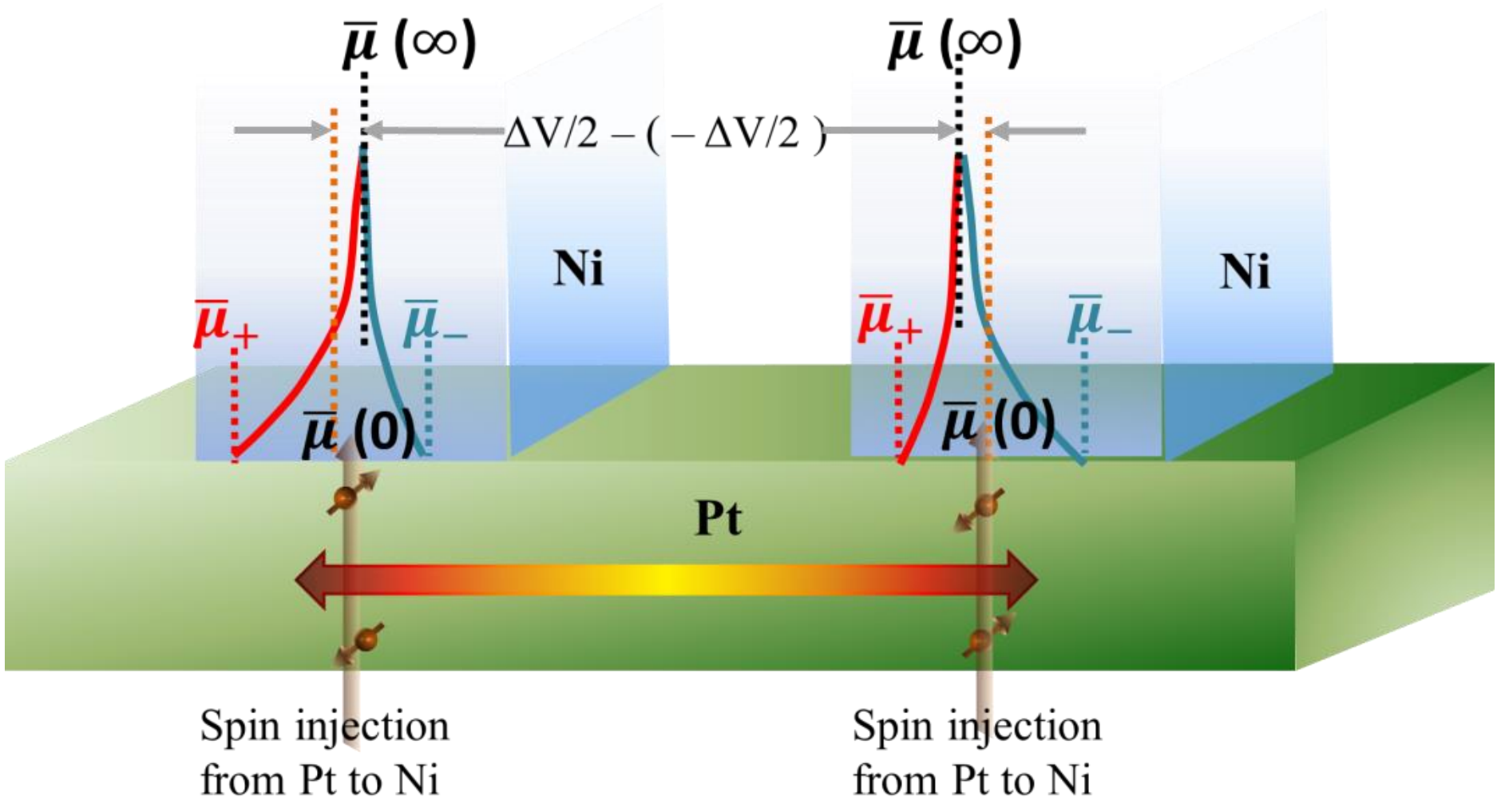

**Figure S2**| Spin injection from Pt to Ni by SNE or SHE.

### S3. More experimental data

### S3.1 Comparison of measured voltage from various terminals

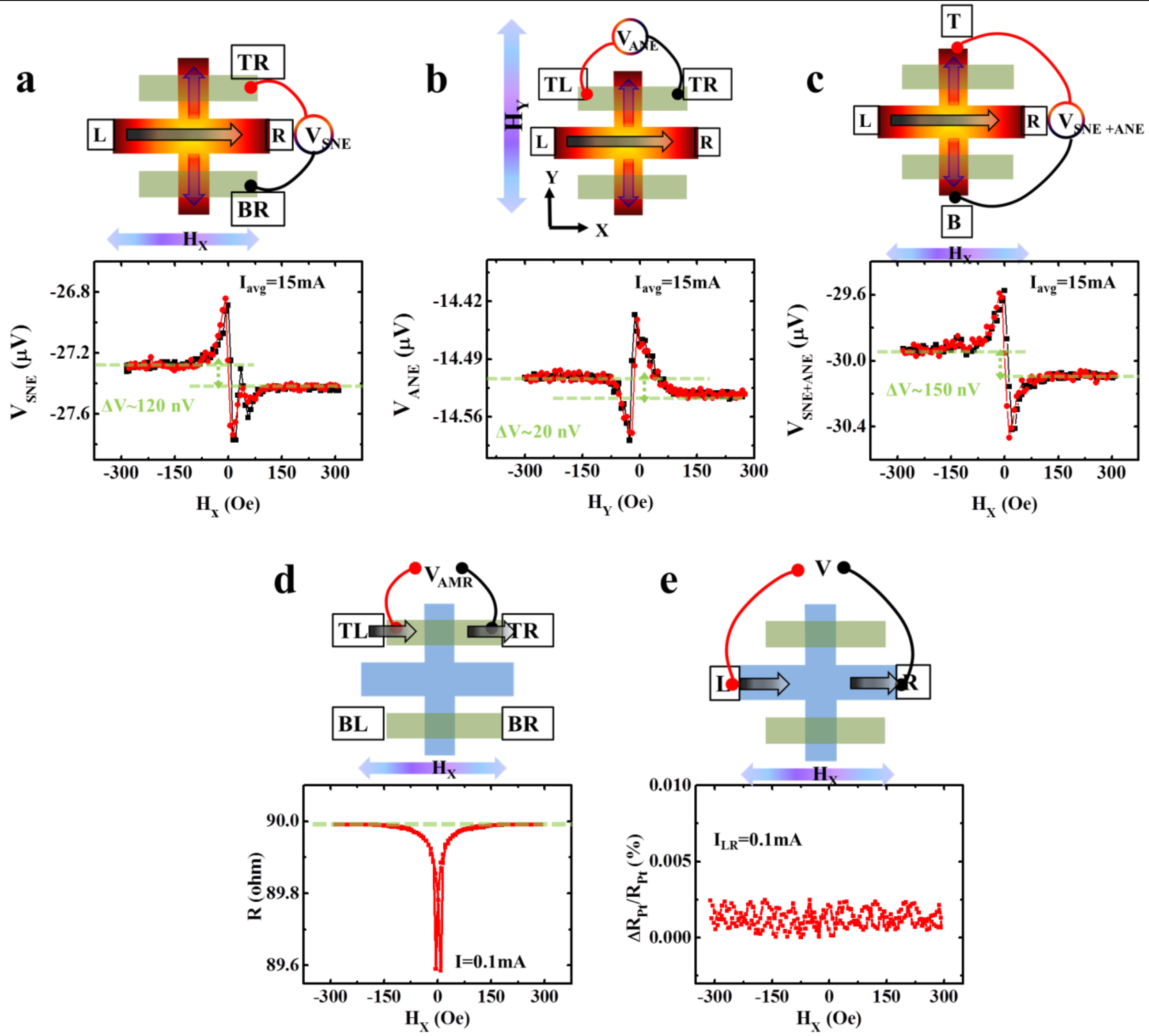

**Figure S3.1**| Comparative study of voltage for various configurations. **a**, $V_{SNE}$ by measuring voltage between two different Ni lines sweeping field along X. **b**, $V_{ANE}$ by measuring voltage between two ends of same Ni terminals sweeping field along Y. **c**, Measured voltage between top and bottom contacts as a function of external field sweep along X. **d**, Conventional AMR measurement of Ni line. **e**, MR measurement of Pt heater line between contact 'L' and 'R'

Figure S3.1 describes complete analysis of experiment including SNE, ANE, AMR and MR measurements. Cross bar shown at centre of all the figures of S3.1.a-e consists of Pt and additional wires show Ni lines. Black (red) arrows indicate the direction of charge (heat) current flow in the system. Fig S3.1.a and S3.1.c are also described in the main article which demonstrates the measurement of SNE and ANE respectively. For 15 mA heater current calculated temperature gradient along Y direction is 8.5 K/μm. While voltage is measured between two different Ni lines ('TR' and 'BR' or 'TL' and 'BL' etc.) sweeping magnetic field along X, we obtain contribution of SNE in form of voltage step (120 nV in fig S3.1.a). Now if we measure voltage between two ends of same Ni terminals ('TR' and 'TL' or 'BR' and 'BL' in fig S3.1.b) sweeping magnetic field along Y, we measure contribution from ANE in form of voltage step since $\nabla V_{ANE}(x)\alpha\ \nabla T_z \times M_Y$ where $\nabla T_z$ is unintentional temperature gradient along Z and $M_Y$ is magnetization of Ni along Y. We can clearly see that ANE voltage step is much smaller than SNE voltage step (Fig S3.1.a-b). Now if voltage is measured between top (‘T’) and bottom (‘B’) contact in presence of external field sweep along X, we expect to see combination of both ANE and SNE signals (Fig S3.1.c). SNE signal should be there since contacts are made on Ni. ANE signal should also be observed because Ni is magnetized along X, unintentional gradient is along Z and finite length of Ni along Y (5μm on top and 5μm at bottom). But magnitude of observed voltage step is almost same for configuration shown in figure S3.1.a and S3.1.c (120 nV to 140 nV). It is because ANE voltage steps are ten times lesser compared to SNE (Fig S3.1.a-b). In configuration shown in figure S3.1.a contact is made on Ni from sideways and hence ANE signal is short to zero (since length along Y becomes short due to Au contacts) which is not the case in the configuration shown in S3.1.c. Additionally we measure resistance of Ni as a function of external field along X axis (Fig S3.1.d) which shows typical behaviour of AMR signal. Figure S3.1.e shows MR of Pt heater line when 0.1 mA is passed. It clearly shows negligible MR within our measurement sensitivity (0.01%). It is also possible that some minor fraction of heater current may spread through Ni line (see S4 section) but it has negligible effect since MR of Pt line as the current spread is very less. If significant amount of heater current flows through Ni due to current spreading we would observe additional AMR signal in figure S3.1.e which would be quite similar to Figure S3.1.d. But we do not observe any such signature (see S4). Hence we can neglect the effect of current spreading around Ni in context of SNE voltage step.

### S3.2 Details of power variation with positive and negative heater current

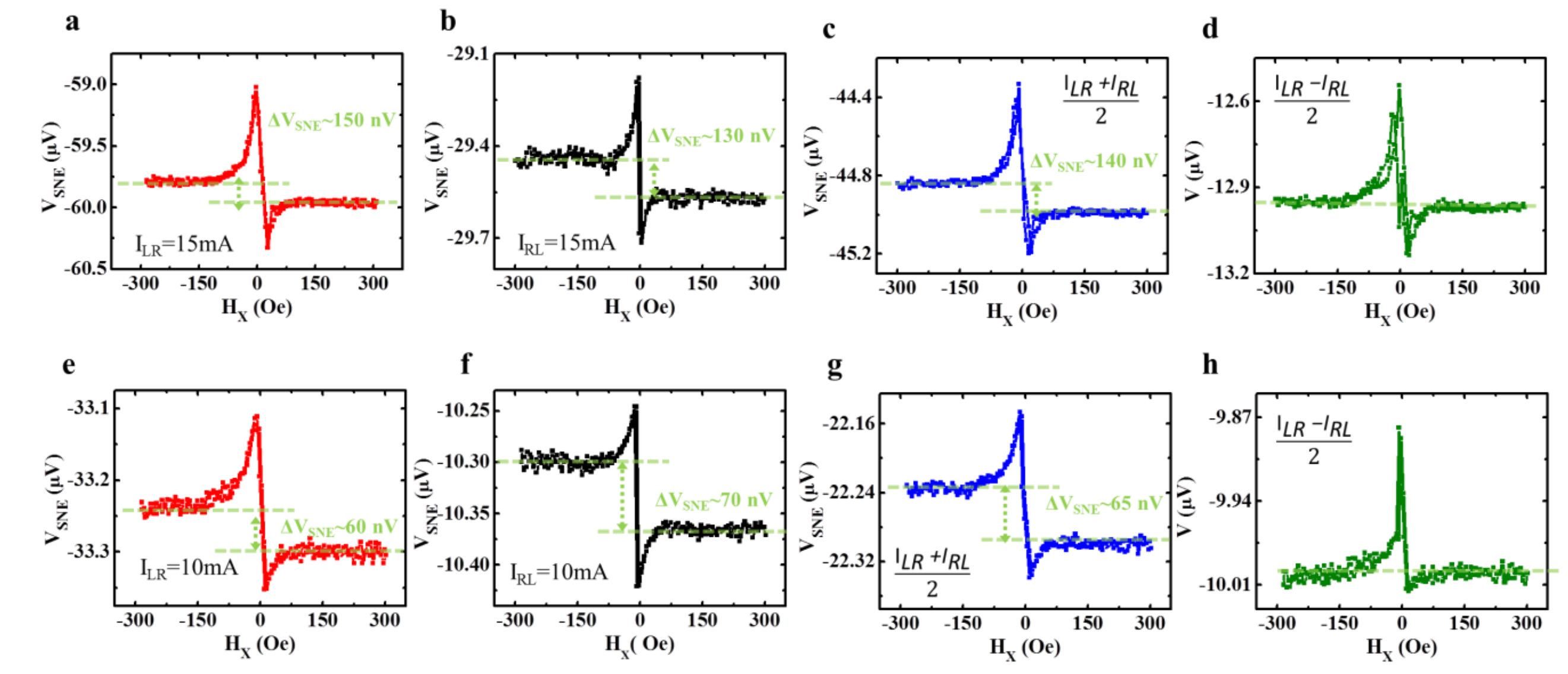

**Figure S3.2**| Detailed analysis of power dependence and heater current polarity alteration. **a-d** show data for ±15 mA heater current. **e-h** show data for ±10 mA heater current.

In this section we shall discuss details of power variation, sample variation and effect of positive and negative heater current (i.e. $I_{LR}$ and $I_{RL}$). We shall focus on the configuration where voltage is measured between two different Ni terminals (for example ‘TR’ & ‘BR’) while field is swept along X. This is typical configuration to measure SNE voltage. Figure S3.2.a,b show the data for +15 mA ($I_{LR}$) and -15 mA ($I_{RL}$) heater current respectively. In both these cases we see signal consists with kinks around centre and a step for higher applied field. Importantly voltage step does not change sign on reversal of heater current but background voltage shifts by some DC values. It clearly indicates that the observed step is originated by heating effect which can be justified by spin Nernst Effect (SNE). Figure S3.2.c is average of positive and negative heater current which gives us sole information of heating effect or SNE. Figure S3.2.a-c look almost similar since heating effect mostly dominates electrical signal. However we cannot rule out the effect from small electrical current leakage at Ni line since background voltage is slightly different for different polarity of applied heater current. On subtraction of S3.2.a and S3.2.b we get contribution of small leakage of electric current. It only contributes to peaks at centre but not to step which is consistent to AMR and PHE (see SI-3.1). Importantly we observe that figure S3.2.c also consists of combination of peaks-dips and step. The step in measured voltage is attributed to SNE whereas kinks are due to thermal AMR and PNE (Planner Nernst effect). Figure S3.2.e-h also show the same for ±10 mA heater current. Since we measure differential voltage between two different Ni lines the background voltage and kinks at centre are taking random values below 100 µV depending upon the asymmetry of device structure. For example in device-2 background voltage for +10 mA applied current is negative (-45.6µV) but for -10 mA current it is +5.3 µV. But convincingly we measure the step in voltage which is deterministic and it does not depend on current polarity. So it is clear indication of SNE voltage step.

| Current (mA) | V(back ground in µV) | V_SNE (nV) | devices |
|---|---|---|---|
| +15 | -60 | 150 | D1 |
| | -83.6 | 115 | D2 |
| | -90.6 | 125 | D3 |
| -15 | -29.5 | 130 | D1 |
| | -10.2 | 120 | D2 |
| | -40.3 | 110 | D3 |
| +10 | -33.2 | 60 | D1 |
| | -45.6 | 52 | D2 |
| | -45.2 | 62 | D3 |
| -10 | -10.3 | 70 | D1 |
| | +5.3 | 55 | D2 |
| | -15.1 | 58 | D3 |

**S4. Effect of current spreading near Ni/Pt bilayer**

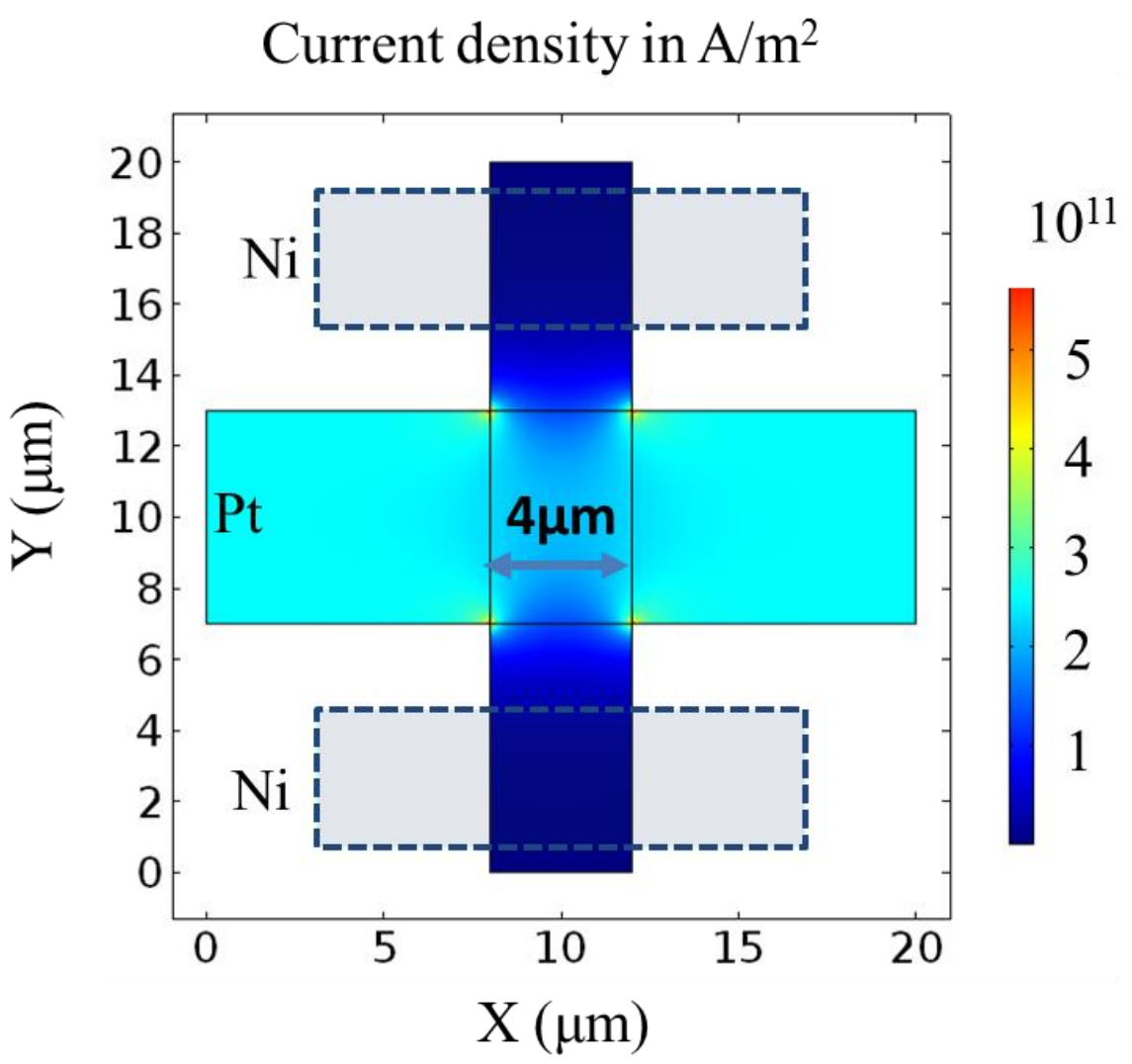

**Figure S2|** Effect of current spreading around Ni contacts

In Pt heater line maximum applied current is 15 mA and corresponding current density is $3x10^{11}$ A/m$^2$. When it passes near the centre of Pt cross bar it can spread towards Ni contact as shown in simulated result of Fig S4. From figure S4 we can estimate that at least 100 times less current flows through Ni line ($1x10^9$ A/m$^2$). Now we have to address whether this minor leakage of current adds any signal to observed SNE voltage. It can contribute in two different ways:

(i) It can add additional electrical signal in measured voltage due to asymmetry of the sample.
(ii) It can cause local heating around Ni to create out of plane gradient.

First of all data shown in figure 3.a, S3.1.a-c and S3.2.c,g are average of positive and negative heater current. So our measured signal is free from any electrical spurious contribution. Secondly, current density under Ni is 100 times smaller compared to actual heater current density. So heating effect will be $10^4$ times lesser which can be negligible. If some unintentional temperature gradient is created in Ni along Z axis that will add to ANE signal which turns out to be very small (Fig S3.1.a and Fig S3.1.b) compared to SNE. Simulated values of current spreading around Ni contacts are consistent with experiments shown in Fig S3.1.e. If significant amount of current leaks through Ni then we could observe magnetoresistance in Pt heater line between terminals 'L' and 'R' itself due to AMR effect of Ni (Fig S3.1.e). But in our experiment magnetoresistance in Pt heater line is not observed (Fig S3.1.e). Additionally in S3.2 section we have individually shown the results while positive and negative heater current is passed (Fig S3.2.a-b.e-f). These results indicate that voltage step is invariant of polarity of heater current which can be only explained by SNE. However there can be additional DC background of measured voltage due to current spreading around Ni and asymmetry of device structure. We can figure out the direct effect of minor leakage of current by subtracting the data of positive heater current and negative heater current. It is shown in figure S3.2.d and S3.2.h. It clearly shows that additional background appears along with small kinks in the signal due to current leakage but we never got any evidence of voltage step due to this (Fig S3.2.d and S3.2.h). Current spreading effect is less since Ni contacts are quite away from centre of the hot spot (5μm).

**S5. Proximity effect and other possible source of voltage step vs SNE signal**

There is a possibility that few monolayers of Pt can become magnetized when it comes in contact of ferromagnet. It is known as magnetic proximity effect [s2]. It is very important issue for YIG/Pt bilayer since YIG is insulator and Pt is conductive. Origin of such observed magnetoresistance in YIG/Pt can be due to magnetized Pt [s2] or it can be simply coexistence of SHE and ISHE [s3-5]. Different groups have ruled out the possibly of proximity effect by rotating magnetic field in three different directions and putting Cu in between YIG and Pt [s3-7].

Our device consists of Pt(10nm)/Ni(10nm) in which both layers are conductive and magnetic property of the system is completely dominated by the magnetization of Ni. Apart from that 10 nm thicker Pt can never be completely magnetized by proximity effect. So we always have heterostructure of heavy metal (HM) and ferromagnet (FM) in which HM will inject spin current and FM will detect. So our justification of SNE behind the observed voltage step in figure 3.a is quite valid despite magnetic proximity effect. Due to this reason magnetic proximity effect is neglected for FM/HM bilayer in previous studies [s8-11].

It is already established that thermopower of magnetic material saturate for higher values of magnetic field since it obeys either sin2θ or cos2θ dependence depending on relative direction of current (heat) and magnetization []. Hence step observed in our experiment between higher values of field (+300Oe and -300 Oe) it has to come from SNE. Simultaneously we observe negligible step when Pt is replaced by Al or field swept normal to accumulated spins (Fig 3.d in main article).